\documentclass[aps,prl,reprint,superscriptaddress]{revtex4-1}

\usepackage{amsmath}
\usepackage[amssymb]{SIunits}
\usepackage{url}
\usepackage{graphicx}
\usepackage{color}

\usepackage{soul}

\usepackage{hyperref}
\usepackage{mathtools}

\usepackage{lipsum}

\usepackage{xcolor}
    \usepackage{changes} 
\usepackage{ulem}
\usepackage{comment}
\definechangesauthor[name={Benedikt}, color={blue}]{be}
\definechangesauthor[name={Johannes}, color={red}]{jo}
\definechangesauthor[name={Dominik}, color={violet}]{do}
\definechangesauthor[name={Stefan}, color={blue}]{st}
\definechangesauthor[name={Stefan}, color={orange}]{stc}

\begin{document}

\title{In-phase and anti-phase synchronization in a laser frequency comb}


\author{Johannes Hillbrand}
\email{johannes.hillbrand@tuwien.ac.at}
\affiliation{Institute of Solid State Electronics, TU Wien, Gusshausstrasse 25-25a, 1040 Vienna, Austria}
\author{Dominik Auth}
\affiliation{Institute of Applied Physics, Technische Universit{\"a}t Darmstadt, Schlossgartenstrasse 7, 64289 Darmstadt, Germany}
\author{Marco Piccardo}
\affiliation{John A. Paulson School of Engineering and Applied Sciences, Harvard University, Cambridge, Massachusetts 02138 USA}
\author{Nikola Opa{\v c}ak}
\author{Gottfried Strasser}
\affiliation{Institute of Solid State Electronics, TU Wien, Gusshausstrasse 25-25a, 1040 Vienna, Austria}
\author{Federico Capasso}
\affiliation{John A. Paulson School of Engineering and Applied Sciences, Harvard University, Cambridge, Massachusetts 02138 USA}
\author{Stefan Breuer}
\affiliation{Institute of Applied Physics, Technische Universit{\"a}t Darmstadt, Schlossgartenstrasse 7, 64289 Darmstadt, Germany}
\author{Benedikt Schwarz}
\email{benedikt.schwarz@tuwien.ac.at}
\affiliation{Institute of Solid State Electronics, TU Wien, Gusshausstrasse 25-25a, 1040 Vienna, Austria}
\affiliation{John A. Paulson School of Engineering and Applied Sciences, Harvard University, Cambridge, Massachusetts 02138 USA}



\begin{abstract}
Coupled clocks are a classic example of a synchronization system leading to periodic collective oscillations. This phenomenon already attracted the attention of Christian Huygens back in 1665, who described it as a kind of ``sympathy" among oscillators. 
In this work we describe the formation of two types of  laser frequency combs as a system of oscillators coupled through the beating of the lasing modes.
We experimentally show two completely different types of synchronizations in a quantum dot laser -- in-phase and splay states. Both states can be generated in the same device, just by varying the damping losses of the system. This effectively modifies the coupling among the oscillators. The temporal output of the laser is characterized using both linear and quadratic autocorrelation techniques. Our results show that both pulses and frequency-modulated states can be generated on demand.
%
%
These findings allow to connect laser frequency combs produced by amplitude-modulated and frequency-modulated lasers, and link these to pattern formation in coupled systems such as Josephson-junction arrays. 

\end{abstract}

\maketitle 

\begin{figure*}
	\centering
	\includegraphics[width=1\linewidth]{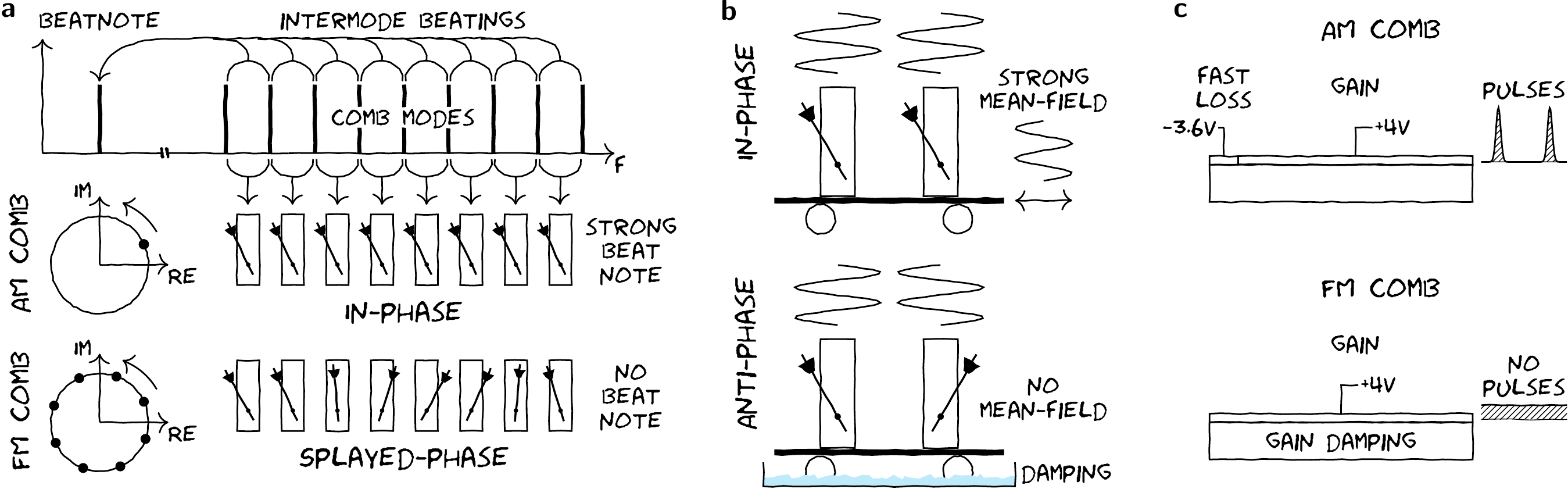}
	\caption{\textbf{a:} Different synchronization states of intermode beatings. Every intermode beating represents an oscillator, here illustrated by metronomes. These oscillators are coupled through their collective action -- the laser beatnote. Depending on the coupling, two distinct synchronization states can be observed. In an AM comb they are synchronized in-phase leading to strong amplitude modulation or pulses. In an FM comb the phases are splayed across the unit circle. Hence, intermode beatings oscillating out-of-phase by $\pi$ mutually annihilate each other. This suppresses amplitude modulations and the beatnote. 
	\textbf{b:} The same can be observed in a system of two coupled metronomes on a movable platform positioned on two cans, where the metronomes will synchronize either in-phase or in anti-phase, depending on the damping. Such damping can be added by placing the cans in a water bath.
	\textbf{c:} Illustrating the analogy to AM and FM comb generation with semiconductor lasers. AM combs can be generated through passive mode-locking via fast saturable loss, while FM comb can even be realized in single section lasers. There, gain saturation dampens amplitude modulations and the beatnote similarly to the water bath in \textbf{b}. 
	}
	\label{fig1}
\end{figure*}

The spectral output of a laser generating an optical frequency comb (OFC) consists of a set of discrete modes that are equidistantly spaced in the frequency domain~\cite{udem2002optical}. In the time domain, the laser produces a periodic output intensity with a temporal waveform that depends on the phase relationship between the modes. For the vast majority of laser frequency combs, such temporal waveform will be predominantly characterized by one of the two following behaviors.
Amplitude-modulated (AM) mode-locked lasers, in which OFCs were first demonstrated, emit a periodic train of light pulses~\cite{keilmann2004time,silberberg1984passive}. 
Their generation relies on mechanisms such as saturable absorption and Kerr lensing. 
On the other hand, frequency-modulated (FM) lasers, which are mostly semiconductor based~\cite{hugi2012mid,schwarz2019monolithic,heck2007observation,rosales2012high,Weber:19} and started emerging more recently, produce a quasi continuous-wave intensity accompanied by a frequency chirp.
FM comb formation has mostly been related to ultra-fast gain dynamics that lead to efficient four-wave mixing~\cite{hugi2012mid,kuehn2008ultrafast,borri2006ultrafast,Bardella:17}, while fundamental self-starting AM comb formation is known to occur in slow gain media with gain relaxation times longer than the cavity round trip time. 
Hence, there is an established belief that AM and FM comb formation exclude each other. 

The formation of a frequency comb can also be understood as a synchronization of coupled oscillators. Each pair of neighboring comb lines produces a beating at their difference frequency. These intermode beatings add up to the so called laser beatnote, as illustrated in Fig.~\ref{fig1}a. This beatnote appears narrow when a comb is formed, indicating the equidistantly spaced comb lines.
Its amplitude depends on the phase relation between the modes.
In-phase synchronization leads to an AM comb with a strong beatnote.
In contrast, anti-phase synchronization generates an FM comb and suppresses the beatnote.
The particular anti-phase state, which was observed in FM combs\cite{singleton2018evidence,schwarz2019monolithic,rosales2012high}, corresponds to splay state synchronization, where the phases are splayed around the unit circle. This results in mutual annihilation of the intermode beatings. 

This interpretation allows us to connect frequency combs to coupled clocks. The first studies of anti-phase synchronization trace back to 1665, when Christiaan Huygens noticed an intriguing phenomenon: two of his newly invented pendulum clocks, mounted on a wooden bar, were swinging in perfect consonance in opposite direction~\cite{huygens1888oeuvres}. He described it as ``an odd kind of sympathy". His observation can be illustrated using coupled metronomes, as sketched in Fig.~\ref{fig1}b. While in this configuration the metronomes tend to synchronize in-phase, the original observation of Huygens can be reproduced by adding a damping that suppresses the movement of the base~\cite{pantaleone2002synchronization}. Here, the damping (or the dissipation) in the system decides, which of the two states will be observed.

In this work, we show that this interpretation of in-phase and anti-phase synchronization can also be applied to semiconductor lasers. We demonstrate experimentally  that both AM and FM comb states can be generated in the very same laser, while leaving the gain dynamics unchanged. Therefore, we can conclude that AM and FM comb generation in semiconductor lasers do not necessarily exclude each other.

%
Semiconductor quantum dot lasers (QDLs) constitute a unique platform for this experimental study. QDLs are semiconductor lasers emitting in the near-infrared with outstanding performance, low threshold current and broad spectral  coverage~\cite{rafailov2007mode,liu2005high,bimberg1997qd}. The active material of these lasers is endowed with properties that can enable both AM and FM combs. On the one hand, the optical gain in QDLs is provided by interband transitions with an upper-state lifetime much longer than the cavity round-trip time. Then, by applying a reverse bias voltage to a short section of the laser, the active region becomes a fast saturable absorber~\cite{rafailov2007mode,thompson2006subpicosecond}. This allows AM comb operation by passive mode-locking ~\cite{haus2000mode,Bardella:18,Meinecke2019} (Fig.~\ref{fig1}c, top).
On the other hand, carrier diffusion is strongly suppressed in these lasers due to carrier localization in the quantum dots~\cite{bardella2017self}. Hence, the standing-wave pattern formed by the longitudinal cavity modes of a Fabry-Perot QDL leads to spatial hole burning. This triggers the multi-mode instability required for FM combs~\cite{tang1963spectral,opacak2019theory}.
FM combs have mostly been observed in single section QDLs and have been referred to as self mode-locking~\cite{heck2007observation,Rosales2012,Weber:19} (Fig.~\ref{fig1}c, bottom).

\begin{figure*}[t]
	\centering
	\includegraphics[width=0.99\linewidth]{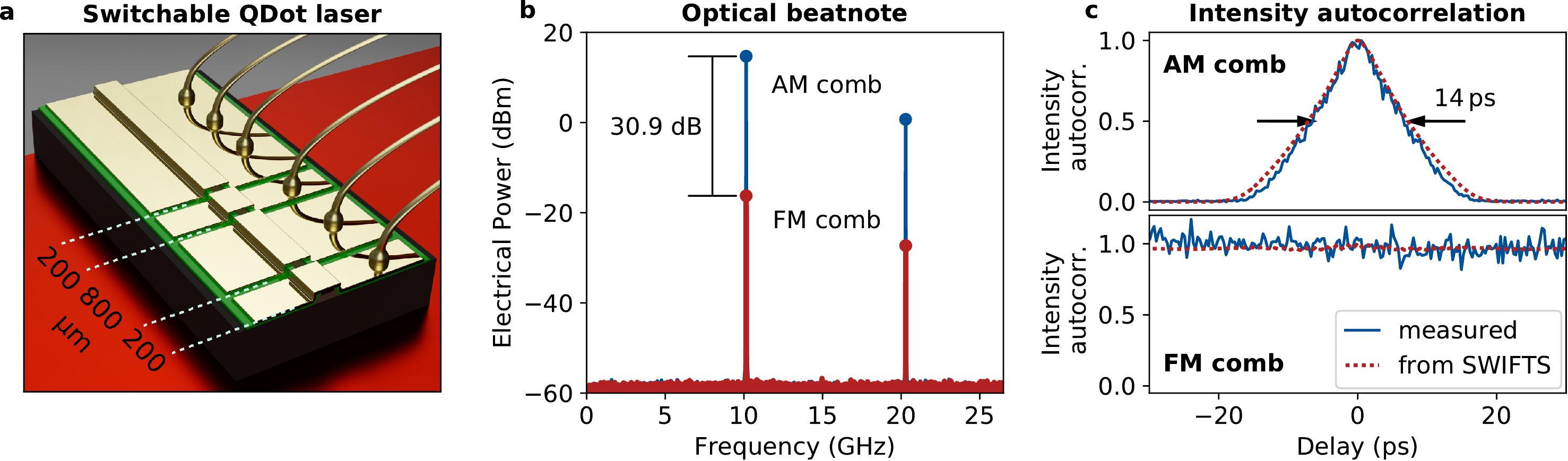}
	\caption{	    \textbf{a:} Schematic of the investigated 4\,mm long QDL with two gain sections and two $200\,\micro\meter$-long absorber sections. \textbf{b:} 
	First and second harmonic beatnote of the QDL, indicating the strong suppression of amplitude modulations in the FM comb regime. The beatnote was measured using a fast photodiode. \textbf{c:} Characterization of the AM and FM state by intensity autocorrelation measurements (blue lines) indicating a width of 14\,ps, corresponding to 10\,ps pulse width assuming a Gaussian shape. In addition, SWIFTS provides a more precise characterization and results for AM and FM comb operation are shown (top, bottom). The red dotted line corresponds to the calculated autocorrelation using the SWIFTS time traces shown in Fig.~\ref{fig3}~a,b and the obtained results highlight the good qualitative agreement between the two experimental techniques.}
	\label{fig2}
\end{figure*}

The investigated laser is a 4\,mm long  monolithic InAs/InGaAs QDL emitting at around 1265\,nm (7900\,cm$^{-1}$). The electrical contacts are divided into two gain sections and two $200\,\micro\meter$-long absorber sections (Fig.~\ref{fig2}a). When all sections are biased homogeneously, the lasing threshold is at 45\,mA with a slope efficieny of 0.22\,mW/mA.
In the following experiments the bias of the gain section is kept constant at 160\,mA to ensure that the gain dynamics remain unchanged. This allows us to prove that AM and FM comb generation do not exclude each other in terms of the gain dynamics. The absorber bias was reconfigured from reverse bias to forward bias in order to identify regions of a narrow beatnote. 
Fig.~\ref{fig2}b shows two beatnote measurement results obtained  with a fast photodiode (electrical bandwidth $>$\,45\,GHz) and for two selected absorber reverse bias voltages of $0\,V$ (red) and $-3.8\,V$ (blue). At $-3.8\,V$ reverse bias, the photogenerated carriers can escape the quantum dots efficiently, resulting in fast saturable losses. There, the laser operates in the AM regime 
exhibiting a narrow and strong beatnote, corresponding to passive mode-locking with fast saturable loss. The intensity autocorrelation measurement results depicted in Fig.~\ref{fig2}c prove that optical pulses are emitted in this regime. 
We retrieve an optical pulse width of 10\,ps from the intensity autocorrelation width of 14\,ps~(FWHM) by assuming a Gaussian shape.

The laser dynamics change drastically, when the fast saturable absorption is switched off by turning off the reverse bias. As expected, the laser no longer generates optical pulses because the gain medium damps amplitude modulations. This is because an amplitude modulated intensity saturates the gain stronger than its continuous average. Hence, a continuous waveform experiences more gain than an AM waveform. Surprisingly, still a narrow beatnote can be observed. This indicates that the QDL is still operating as a frequency comb in absence of fast saturable losses. The suppression of amplitude modulations is evident in the strong attenuation of the beatnote by 31~dB (Fig.~\ref{fig2}b), while the total output power even increases from $15.6\,\milli\watt$ to $22.3\,\milli\watt$. The reason for this is that by switching off the fast saturable loss, the laser switches from AM to FM comb operation.
The corresponding intensity autocorrelation measurement results in Fig.~\ref{fig2}c confirms the expected flat and continuous-wave laser output, connected to the suppressed beatnote. However, it also shows that intensity autocorrelation yields useful information only for AM combs. 

\begin{figure*}[t]
	\centering
	\includegraphics[width=0.99\linewidth]{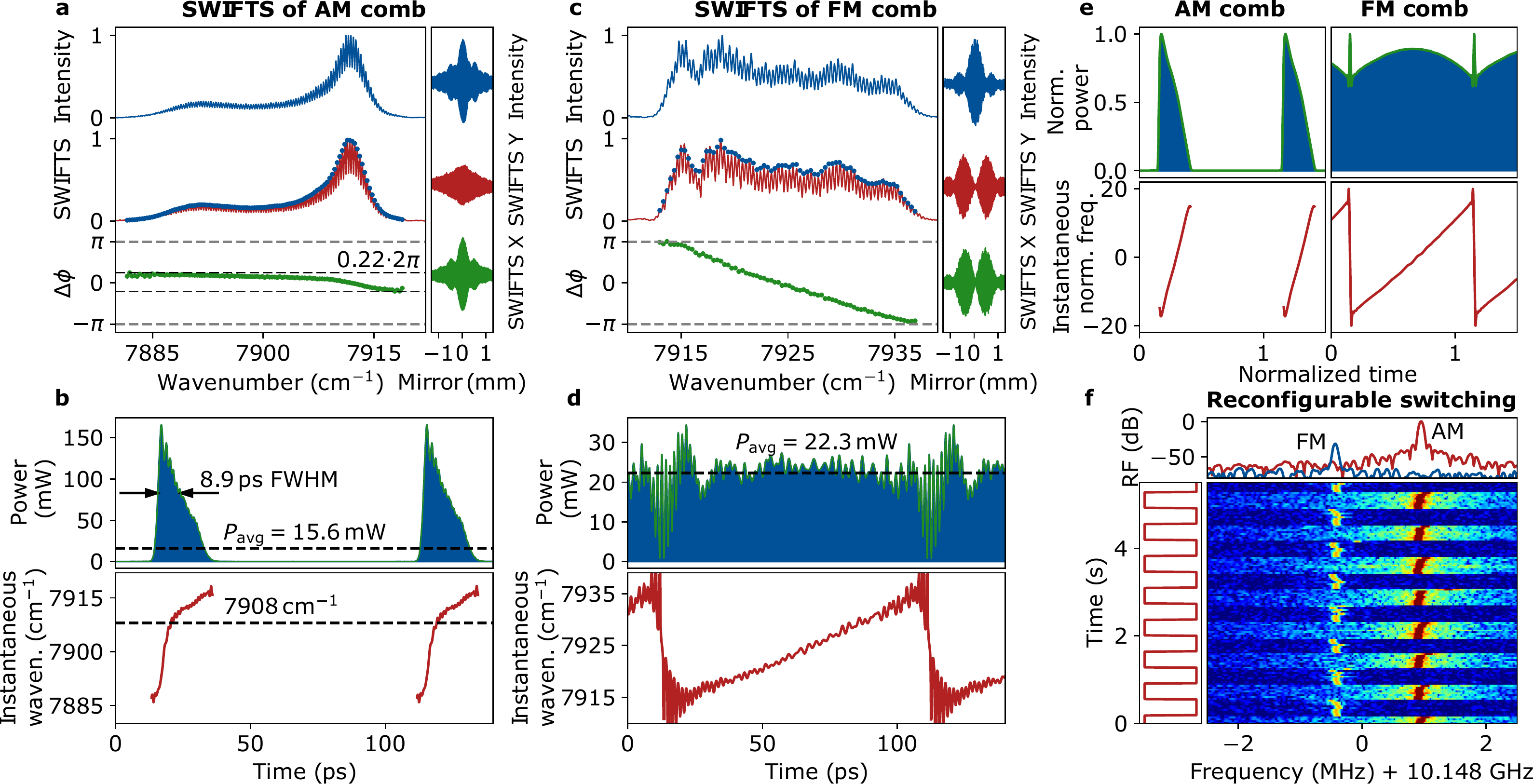}
	\caption{\textbf{a:} SWIFTS characterization of the AM comb state.
	The SWIFTS spectrum (red) proves that the entire intensity spectrum (blue) is locked. The intermodal difference phases $\Delta \phi$ (green dots) cover a range of $0.43\, \pi$. Inset on the right: intensity and SWIFTS interferograms around zero delay. \textbf{b:} The reconstructed time signal of the laser output yields optical pulses with width of 8.9 ps (FWHM).
	\textbf{c:} SWIFTS characterization of the FM QDL comb. In contrast to the passively mode-locked state, the intermodal difference phases are now splayed across the full range of $2\pi$. \textbf{d:} The reconstructed time-domain signal of the FM comb state nearly depicts negligible amplitude modulation while the instantaneous frequency is linearly chirped across one round-trip period. 
	\textbf{e:} Numerical simulations results, depicting the normalized power and instantaneous frequency in the case of both AM and FM combs. The instantaneous frequency and time are normalized to the cavity round-trip time.
	\textbf{f:} Color-coded radiofrequency (RF) intensity in dependence on a rectangular modulation of the absorber section voltage between 0\,V and -3.8\,V showing the switching between the AM and FM comb state. 
	}
	\label{fig3}
\end{figure*}

Shifted-wave interference Fourier transform spectroscopy (SWIFTS) based on a Fourier transform infrared (FTIR) spectrometer is an ideal measurement technique that allows to study both AM and FM combs ~\cite{burghoff2015evaluating}. SWIFTS allows to measure the spectrally-resolved coherence and phases of the intermode beatings. In analogy to synchronization, as depicted in Fig.~\ref{fig1}a, it provides access to the periodic oscillation of each metronome, whereas the laser beatnote only describes their collective action. This knowledge provides full insight into the comb operation by enabling the direct reconstruction of the laser output intensity and instantaneous frequency. 
Fig.~\ref{fig3}a shows the experimental SWIFTS results of the QDL with reverse biased absorber -3.8\,V. The spectrum of the QDL spans across 30\,cm$^{-1}$ and consists of a group of modes centered at 7908\,cm$^{-1}$ and a long tail towards lower wave numbers. The SWIFTS spectrum has the same shape as the intensity spectrum without any spectral holes, proving frequency comb operation across the entire lasing spectrum. The intermode beating phases $\Delta \phi$ are dominantly synchronized in-phase with a small remaining chirp at the high energy end of the spectrum. Consequently, they span across a narrow range of 0.43$\pi$. $\Delta \phi$ is related to the spectral group delay with $2\pi$ corresponding to exactly one cavity round-trip period. Hence, we expect non-transform limited optical pulses with a duty cycle of $0.43\pi/2\pi=21.5\%$. 
A second clear indicator for strong amplitude modulation are the local maxima of the SWIFTS interferograms at zero path difference of the FTIR spectrometer (Fig.~\ref{fig3}a).

Indeed, the reconstructed time domain signal of the QDL depicted in Fig. \ref{fig3}b confirms a train of isolated optical pulses with a 8.9\,ps full width at half maximum (FWHM). The peak power is 167\,mW corresponding to an enhancement of 10.7 with respect to the average power of 15.6\,mW. The pulses show an intense initial burst followed by a trailing edge. In order to independently verify the SWIFTS results, we show the excellent agreement of the time-domain reconstruction retrieved from SWIFTS with the measured nonlinear intensity autocorrelation in Fig.~\ref{fig2}c. The reconstructed instantaneous wavenumber (Fig.~\ref{fig3}b) allows to identify the spectral regions responsible for this pulse shape. While the intense initial burst contains the wavenumbers up to 7908\,cm$^{-1}$ (tail of the spectrum in Fig. \ref{fig3}a), the trailing edge of the pulse is caused by wavenumbers above 7908\,cm$^{-1}$. In fact, already the intermodal difference phases in Fig. \ref{fig3}a indicate that this spectral region is strongly chirped.


The experimental results of the SWIFTS characterization of the FM state is shown in Fig.~\ref{fig3}c. The entire emission spectrum is phase-locked. In contrast to the AM comb state (Fig.~\ref{fig3}a), both SWIFTS interferograms have a local minimum at zero delay which can be linked to the suppression of amplitude modulation~\cite{hugi2012mid}. Even more intriguing is the obtained linear phase pattern that corresponds to splay phase synchronization. The corresponding time domain signal (Fig. \ref{fig3}c top) shows almost no amplitude modulation and a linearly chirped instantaneous frequency. This is in good agreement with previous experimental observations of a strong chirp in QDL~\cite{heck2007observation}. The same phenomenon was observed in other lasers, such as quantum cascade lasers and interband cascade lasers~\cite{singleton2018evidence,hillbrand2018coherent,schwarz2019monolithic}.

In Fig.~\ref{fig3}e we show the results of numerical simulations that are based on the time domain traveling model. The experimental time traces for the QDL output power and instantaneous frequency, given on Figs.~\ref{fig3}b and d, are recreated. In the case of an FM state, where the device is uniformly biased (Fig. 3e right), a continuous laser output is obtained by gain dampening of the amplitude modulations. The characteristic chirped frequency is particularly a result of interplay between finite group velocity dispersion (GVD) and Kerr nonlinearity, as was revealed in recent theoretical work \cite{opacak2019theory}. As the GVD contribution is independent from the time dynamics of the gain medium, this provides further evidence that FM combs are not limited to ultra-fast lasers. On the other hand, we create perfect conditions for passive mode-locking by applying a reverse bias to a short section of the cavity, as well as lowering the saturation intensity and drastically decreasing the carrier lifetime in this section. As a result, chirped pulses are obtained, as is shown (Fig.3e left). A trailing edge similar to the experimentally observed one is clearly visible.



The damping mechanism that chooses the comb state is due to the gain medium itself. This is because a modulated intensity saturates the gain stronger than its continuous average. As this gain damping is always present, switching between AM and FM states is here performed by introducing the reversed effect through saturable losses. In order to show that this switching can be achieved in a deterministic and temporally reconfigurable way, we apply a periodic rectangular voltage modulation to the absorber section between 0\,V and -3.8\,V. The temporal evolution of the beatnote is plotted in Fig.~\ref{fig3}f. Indeed, the QDL switches back and forth between the AM and the FM state by periodically changing the damping mechanism.

In conclusion, we demonstrated AM and FM comb operation - two distinct frequency comb states with entirely different temporal dynamics - in a single semiconductor quantum dot laser by changing the absorber section bias. Their underlying physical mechanisms can be directly related to oscillators coupled through the beating of the lasing modes. 
%
By both non-linear intensity autocorrelation temporal domain and beat note spectroscopy analysis, we identified optical pulse shape and width as well as the specific spectral regions responsible for pulse broadening. 
%
%
%
%
%
%
We numerically and experimentally highlighted that the requirements on the gain dynamics for AM and FM combs do not necessarily exclude each other and that FM combs can be generated also with slower gain media where the gain relaxation is slower than the round-trip time, when spatial hole burning is strong enough to trigger multi-mode operation and sufficient group velocity dispersion is present. 
%

This work was supported by the Austrian Science Fund (FWF) within the projects ”NanoPlas” (P28914-N27) and "Building Solids for Function" (Project W1243), the city of Vienna within the "Hochschuljubil{\"a}umsstiftung", the German Research Foundation (DFG) (389193326) and the Adolf Messer Foundation (doctoral fellowship). M.P. and F.C. acknowledge support from the National Science Foundation under Award No. CCSS-1807323. Any opinions, findings, conclusions or recommendations expressed in this material are those of the authors and do not necessarily reflect the views of the National Science Foundation.

\bibliography{literature.bib}


\end{document}